\documentclass[conference, 10pt]{IEEEtran}

\usepackage[caption=false,font=normalsize,labelfont=sf,textfont=sf]{subfig}
\usepackage{graphicx}
\usepackage{color}
\usepackage{placeins}
\usepackage{float}
\usepackage{tabularx,colortbl}
\usepackage[nolist]{acronym}

\begin{document}

\begin{acronym}

\acro{2D}{Two Dimensions}%
\acro{2G}{Second Generation}%
\acro{3D}{Three Dimensions}%
\acro{3G}{Third Generation}%
\acro{3GPP}{Third Generation Partnership Project}%
\acro{3GPP2}{Third Generation Partnership Project 2}%
\acro{4G}{Fourth Generation}%
\acro{5G}{Fifth Generation}%

\acro{AI}{Artificial Intelligence}%
\acro{AoA}{Angle of Arrival}%
\acro{AoD}{Angle of Departure}%
\acro{AR}{Augmented Reality}%
\acro{AP}{Access Point}
\acro{AE}{Antenna Element}
\acro{AC}{Anechoic Chamber}
\acro{AUT}{Antenna Under Test}
\acro{AD}{Anderson-Darling}%
\acro{ANOVA}{Analysis Of Variance}%

\acro{BER}{Bit Error Rate}%
\acro{BPSK}{Binary Phase-Shift Keying}%
\acro{BRDF}{ Bidirectional Reflectance Distribution Function}%
\acro{BS}{Base Station}%

\acro{CA}{Carrier Aggregation}%
\acro{CDF}{Cumulative Distribution Function}%
\acro{CDM}{Code Division Multiplexing}%
\acro{CDMA}{Code Division Multiple Access}%
\acro{CPU} {Central Processing Unit}
\acro{CUDA}{Compute Unified Device Architecture}
\acro{CDF}{Cumulative Distribution Function}
\acro{CI}{Confidence Interval}
\acro{CVRP}{Constrained-View Radiated Power}
\acro{CATR}{Compact Antenna Test Range}
\acro{CV}{Coefficient of Variation}
\acro{CTIA}{Cellular Telephone Industries Association}
 
\acro{D2D}{Device-to-Device}%
\acro{DL}{Down Link}%
\acro{DS}{Delay Spread}%
\acro{DAS}{Distributed Antenna System}
\acro{DKED}{double knife-edge diffraction}
\acro{DUT}{Device Under Test}
\acro{DR}{Dynamic Range}


\acro{EDGE}{Enhanced Data rates for GSM Evolution}%
\acro{EIRP}{Equivalent Isotropic Radiated Power}%
\acro{eMBB}{Enhanced Mobile Broadband}%
\acro{eNodeB}{evolved Node B}%
\acro{ETSI}{European Telecommunications Standards Institute}%
\acro{ER}{Effective Roughness}%
\acro{E-UTRA}{Evolved UMTS Terrestrial Radio Access}%
\acro{E-UTRAN}{Evolved UMTS Terrestrial Radio Access Network}%
\acro{EF}{Electric Field}
\acro{EMC}{Electromagnetic Compatibility}

\acro{FDD}{Frequency Division Duplexing}%
\acro{FDM}{Frequency Division Multiplexing}%
\acro{FDMA}{Frequency Division Multiple Access}%
\acro{FoM}{Figure of Merit}
\acro{FoV}{Field of View}
\acro{FSA}{Frequency Selective Absorber}
\acro{FS}{Frequency Samples}
\acro{GI}{Global Illumination} %
\acro{GIS}{Geographic Information System}%
\acro{GO}{Geometrical Optics} %
\acro{GPU}{Graphics Processing Unit}%
\acro{GPGPU}{General Purpose Graphics Processing Unit}%
\acro{GPRS}{General Packet Radio Service}%
\acro{GSM}{Global System for Mobile Communication}%
\acro{GNSS}{Global Navigation Satellite System}%
\acro{GoF}{Goodness-of-Fit}
\acro{H2D}{Human-to-Device}%
\acro{H2H}{Human-to-Human}%
\acro{HDRP}{High Definition Render Pipeline}
\acro{HSDPA}{High Speed Downlink Packet Access}
\acro{HSPA}{High Speed Packet Access}%
\acro{HSPA+}{High Speed Packet Access Evolution}%
\acro{HSUPA}{High Speed Uplink Packet Access}
\acro{HPBW}{Half-Power Beamwidth}
\acro{HA}{Horn Antenna}

\acro{IEEE}{Institute of Electrical and Electronic Engineers}%
\acro{InH}{Indoor Hotspot} %
\acro{IMT} {International Mobile Telecommunications}%
\acro{IMT-2000}{\ac{IMT} 2000}%
\acro{IMT-2020}{\ac{IMT} 2020}%
\acro{IMT-Advanced}{\ac{IMT} Advanced}%
\acro{IoT}{Internet of Things}%
\acro{IP}{Internet Protocol}%
\acro{ITU}{International Telecommunications Union}%
\acro{ITU-R}{\ac{ITU} Radiocommunications Sector}%
\acro{IS-95}{Interim Standard 95}%
\acro{IES}{Inter-Element Spacing}
\acro{IF}{Intermediate Frequency}


\acro{KPI}{Key Performance Indicator}%
\acro{K-S}{Kolmogorov-Smirnov}

\acro{LB} {Light Bounce}
\acro{LIM}{Light Intensity Model}%
\acro{LOS}{Line-Of-Sight}%
\acro{LTE}{Long Term Evolution}%
\acro{LTE-Advanced}{\ac{LTE} Advanced}%
\acro{LSCP}{Lean System Control Plane}%
\acro{LSI} {Light Source Intensity}

\acro{M2M}{Machine-to-Machine}%
\acro{MatSIM}{Multi Agent Transport Simulation}
\acro{METIS}{Mobile and wireless communications Enablers for Twenty-twenty Information Society}%
\acro{METIS-II}{Mobile and wireless communications Enablers for Twenty-twenty Information Society II}%
\acro{MIMO}{Mul\-ti\-ple-In\-put Mul\-ti\-ple-Out\-put}
\acro{mMIMO}{massive MIMO}%
\acro{mMTC}{massive Machine Type Communications}%
\acro{mmW}{millimeter-wave}%
\acro{MU-MIMO}{Multi-User MIMO}
\acro{MMF}{Max-Min Fairness}
\acro{MKED}{Multiple Knife-Edge Diffraction}
\acro{MF}{Matched Filter}
\acro{mmWave}{Millimeter Wave}

\acro{NFV}{Network Functions Virtualization}%
\acro{NLOS}{Non-Line-Of-Sight}%
\acro{NR}{New Radio}%
\acro{NRT}{Non Real Time}%
\acro{NYU}{New York University}%
\acro{N75PRP}{Near-75-degrees Partial Radiated Power}%
\acro{NHPRP}{Near-Horizon Partial Radiated Power}%

\acro{O2I}{Outdoor to Indoor}%
\acro{O2O}{Outdoor to Outdoor}%
\acro{OFDM}{Orthogonal Frequency Division Multiplexing}%
\acro{OFDMA}{Or\-tho\-go\-nal Fre\-quen\-cy Di\-vi\-sion Mul\-ti\-ple Access}
\acro{OtoI}{Outdoor to Indoor}%
\acro{OTA}{Over-The-Air}

\acro{PDF}{Probability Distribution Function}
\acro{PDP}{Power Delay Profile}
\acro{PHY}{Physical}%
\acro{PLE}{Path Loss Exponent}
\acro{PRP}{Partial Radiated Power}
\acro{PW}{Plane Wave}%
\acro{PR}{Pass Rate}%
\acro{PAS}{Power-Angle Spectrum}

\acro{QAM}{Quadrature Amplitude Modulation}%
\acro{QoS}{Quality of Service}%

\acro{RCSP}{Receive Signal Code Power}
\acro{RAN}{Radio Access Network}%
\acro{RAT}{Radio Access Technology}%

\acro{RAN}{Radio Access Network}%
\acro{RMa}{Rural Macro-cell}%
\acro{RMSE} {Root Mean Square Error}
\acro{RSCP}{Receive Signal Code Power}%
\acro{RT}{Ray Tracing}
\acro{RX}{receiver}
\acro{RMS}{Root Mean Square}
\acro{Random-LOS}{Random Line-Of-Sight}
\acro{RF}{Radio Frequency}
\acro{RC}{Reverberation Chamber}
\acro{RC-HARC}{Resonating Cavity Hybrid Anechoic-Reverberation Chamber}%
\acro{RIMP}{Rich Isotropic Multipath}
\acro{RHA}{Reference Horn Antenna}
\acro{RIMPMA}{RIMP Measurement Antenna}

\acro{SB} {Shadow Bias}
\acro{SC}{small cell}
\acro{SDN}{Software-Defined Networking}%
\acro{SGE}{Serious Game Engineering}%
\acro{SF}{Shadow Fading}%
\acro{SIMO}{Single Input Multiple Output}%
\acro{SINR}{Signal to Interference plus Noise Ratio}
\acro{SISO}{Single Input Single Output}%
\acro{SMa}{Suburban Macro-cell}%
\acro{SNR}{Signal to Noise Ratio}
\acro{SU}{Single User}%
\acro{SUMO}{Simulation of Urban Mobility}
\acro{SS} {Shadow Strength}
\acro{STD}{Standard Deviation}
\acro{SW} {Sliding Window}


\acro{TDD}{Time Division Duplexing}%
\acro{TDM}{Time Division Multiplexing}%
\acro{TD-CDMA}{Time Division Code Division Multiple Access}%
\acro{TDMA}{Time Division Multiple Access}%
\acro{TX}{transmitter}
\acro{TZ}{Test Zone}
\acro{TRP}{Total Radiated Power}


\acro{UAV}{Unmanned Aerial Vehicle}%
\acro{UE}{User Equipment}%
\acro{UI}{User Interface}
\acro{UHD}{Ultra High Definition}
\acro{UL}{Uplink}%
\acro{UMa}{Urban Macro-cell}%
\acro{UMi}{Urban Micro-cell}%
\acro{uMTC}{ultra-reliable Machine Type Communications}%
\acro{UMTS}{Universal Mobile Telecommunications System}%
\acro{UPM}{Unity Package Manager}
\acro{UTD}{Uniform Theory of Diffraction} %
\acro{UTRA}{{UMTS} Terrestrial Radio Access}%
\acro{UTRAN}{{UMTS} Terrestrial Radio Access Network}%
\acro{URLLC}{Ultra-Reliable and Low Latency Communications}%
\acro{UHRP}{Upper Hemisphere Radiated Power}%
\acro{ULA}{Uniform Linear Array}%

\acro{V2V}{Vehicle-to-Vehicle}%
\acro{V2X}{Vehicle-to-Everything}%
\acro{VP}{Visualization Platform}%
\acro{VR}{Virtual Reality}%
\acro{VNA}{Vector Network Analyzer}
\acro{VIL}{Vehicle-in-the-loop}
\acro{VIRC}{Vibrating Intrinsic Reverberation Chamber}

\acro{WCDMA}{Wideband Code Division Multiple Access}%
\acro{WINNER}{Wireless World Initiative New Radio}%
\acro{WINNER+}{Wireless World Initiative New Radio +}%
\acro{WiMAX}{Worldwide Interoperability for Microwave Access}%
\acro{WRC}{World Radiocommunication Conference}%

\acro{xMBB}{extreme Mobile Broadband}%

\acro{ZF}{Zero Forcing}

\end{acronym}
%
\title{RC Measurement Uncertainty Estimation Method for Directive Antennas and Turntable Stirring}

%

\author{\IEEEauthorblockN{Alejandro Antón Ruiz}
\IEEEauthorblockA{\textit{Department of Electrical Engineering} \\
\textit{University of Twente}\\
Enschede, Netherlands \\
a.antonruiz@utwente.nl}
\and
\IEEEauthorblockN{John Kvarnstrand, Klas Arvidsson}
\IEEEauthorblockA{\textit{Bluetest AB}\\
Gothenburg, Sweden \\
name.familyname@bluetest.se}
\and
\IEEEauthorblockN{Andrés Alayón Glazunov}
\IEEEauthorblockA{\textit{Department of Science and Technology} \\
\textit{Linköping University}\\
Norrköping Campus, Sweden \\
andres.alayon.glazunov@liu.se}
}

\maketitle
\begin{abstract}
This paper investigates measurement uncertainty in a Reverberation Chamber (RC) within the lower FR2 bands (24.25-29.5 GHz). The study focuses on the impact of several factors contributing to RC measurement uncertainty, including finite sample size, polarization imbalance, and spatial non-uniformity. A series of 24 measurements were conducted using a horn antenna, known for its directivity in mmWave frequencies, varying antenna parameters such as height, orientation, position on the turntable, and polarization within a predefined chamber volume. The measurement uncertainty was evaluated by a method based on the standardized 3GPP and CTIA approaches, incorporating uncorrelated measurements and analyzing Pearson correlation coefficients between measurement pairs. An analysis of variance (ANOVA) was performed on the frequency-averaged power transfer function to identify the significance and impact of each variable on measurement variability. Additionally, the K-factor was estimated for each measurement set as part of the RC characterization, using an alternative approach to account for the turntable stirring effect. The findings highlight which variables most significantly influence measurement uncertainty, where the antenna orientation emerges as the most significant
factor for the mmWave directive antenna setup.

\end{abstract}

\IEEEpeerreviewmaketitle

\section{Introduction}
\label{S1}
Antenna and \ac{OTA} measurements are crucial to ensure that antennas and devices meet design requirements and comply with regulations. Traditionally, these measurements occur in \acp{AC}, which emulate a Pure-\ac{LOS} environment, allowing directional performance evaluation over a sphere for parameters like \ac{EIRP}. This is particularly useful in applications such as automotive communication, where partial quantities like \ac{PRP} \cite{5GAA,ILMVG} or weighted partial radiation efficiency \cite{WPRE} are of interest. Full-sphere data acquisition enables the computation of integral quantities such as \ac{TRP}.

Alternatively, \ac{OTA} and antenna measurements can be performed in \acp{RC}, which emulate a \ac{RIMP} environment characterized by a Rayleigh distribution in the absence of a \ac{LOS} component \cite{LTEBOOK}. \acp{RC} offer faster acquisition of integral quantities like antenna efficiency and \ac{TRP} compared to anechoic chambers \cite{BluetestCATR}, and can also support \ac{MIMO} measurements by emulating scenarios with appropriate delay spread and i.i.d. Rayleigh fading channels \cite{OTAMIMO}.

\ac{mmWave}, a key enabler of \ac{5G} \cite{mmWavejust}, provides increased capacity due to its large bandwidth. However, \ac{mmWave} channels are more directional with higher path loss, requiring directive antennas capable of beamforming. While directional data, such as radiation patterns and partial quantities like \ac{PRP}, is crucial, integral quantities remain essential for devices like mobile phones, where orientation is not fixed \cite{LTEBOOK}, for compliance testing.

In this paper, we focus on traditional \ac{RC} operation for \ac{mmWave} systems, emphasizing the uncertainty analysis due to spatial non-uniformity, finite sampling, and polarization imbalance with turntable stirring \cite{TTstirring_8}. We use a standardized method that involves repeated measurements at different positions within the test volume, estimating uncertainty from the standard deviation. Independent measurements are verified using Pearson correlation coefficients. An \ac{ANOVA} assesses the significance of variables—height, antenna orientation, turntable position, and polarization—on the frequency-averaged power transfer function, prioritizing variables for optimal uncertainty capture under time constraints.

We also propose a method for estimating the $K$-factor in \acp{RC} with turntables, which is critical for evaluating deviation from the ideal Rayleigh environment and impacts uncertainty. Using a directive horn antenna, common in \ac{mmWave} applications, we observe the relationship between $K$-factor and uncertainty, as suggested in the literature, though a detailed comparison is left for future work. The proposed $K$-factor estimation method, while based on existing components, is novel in its joint application and demonstrates superior performance in Monte Carlo simulations.

\section{Considered uncertainty components}
\label{UCComps}
The uncertainty of the power transfer function $G_{ref}$ is assessed as a key \ac{RC} measurement parameter because it contributes to the uncertainty evaluation of other measurements and is used for calibration reference. 

\subsection{Lack of spatial uniformity}
\label{LOSU}
In an \ac{RC}, the \ac{RIMP} environment implies that a large number of plane waves impinge into the \ac{DUT}. Independency in polarization, amplitude, phase, and \acp{AoA} is a requirement. The \ac{LOS} component must be removed leading to the magnitude of the complex signal becoming Rayleigh distributed, the phase uniformly distributed over $2\pi$, and the power exponentially distributed \cite{LTEBOOK}.

If a \ac{LOS} component is present, then the magnitude of the complex signal will follow the Rician distribution. The Rayleigh distribution is a special case of the Rician distribution where the $K$-factor (the ratio between the \ac{LOS} and \ac{NLOS} components) is 0. Therefore, any deviation from $K=0$ or presence of a \ac{LOS} component will make the environment deviate from the ideal \ac{RIMP}. Thus, the \ac{AoA} distribution is no longer uniform. This has a serious implication given that the \ac{RC} measurement method relies on the statistical averaging of several samples to obtain the integral magnitude of interest, such as \ac{TRP} or antenna efficiency. It is worthwhile to note that $K>0$ can be caused by the \ac{LOS}, but also by other factors, such as chamber loading.

\subsection{Finite number of samples}
\acp{RC} operate on statistical averaging to obtain the quantity of interest, e.g., antenna efficiency or \ac{TRP}. Therefore, the power transfer function of the chamber or $G_{ref}$ is at the focus of our investigation of the measurement uncertainty estimation of magnitudes of interest. The uncertainty of this estimation follows the Gaussian standard deviation, which depends on the number of (effectively) independent samples $N_{eff}$ as $\sigma = 1/N_{eff}$ \cite{GaussianAccGref}. This applies in the absence of a \ac{LOS} contribution \cite{KFUCF}.
The fact that the uncertainty of the estimation of the magnitude of interest inversely depends on $N_{eff}$ is critical to obtaining a sufficiently large $N_{eff}$ to achieve the desired uncertainty. The turntable stirring is key to obtaining a large $N_{eff}$ in some cases, as shown in \cite{TTstirringorig}. This could be assimilated to having an electrically larger chamber, so it allows having smaller chambers while still being able to collect a large $N_{eff}$, thus having less measurement uncertainty of the magnitude of interest. Since most \acp{RC} are designed to measure at sub-$6$~GHz frequencies (usually including sub-$1$~GHz frequencies), it should not be a problem to get a large enough $N_{eff}$ at \ac{mmWave} even without turntable stirring, unless the uncertainty due to the finite number of samples is desired to be kept low. We use turntable stirring at \ac{mmWave}, a common practice in \ac{RC} measurements. Future work will assess whether its benefits, such as averaging the unstirred component \cite{TTstirringorig}, outweigh the increased measurement time.

\subsection{Polarization imbalance}
The large number of plane waves generated in an \acp{RC} ought to be independent in polarization, and maintain polarization balance, i.e., the received power at any given point inside the testing volume should be the same irrespective of the polarization of the receiving antenna. Deviation from this will contribute to measurement uncertainty, making the power transfer function $G_{ref}$ depend on the \ac{DUT} polarization.

\section{Uncertainty characterization}
\subsection{Chamber precharacterization measurements}
\ac{3GPP} \cite{3GPPRCStandard} and \ac{CTIA} \cite{CTIAprechar}  propose a characterization of the transfer function uniformity and uncertainty due to spatial non-uniformity, respectively. These procedures precede \ac{DUT} measurements, hence the term "chamber precharacterization measurements". In both cases, this is evaluated as the estimated standard deviation from a series of measurements of the power transfer function $G_{ref}$  using a reference antenna. According to \ac{3GPP}, the measurements shall be taken at least at $3$ uncorrelated locations and $6$ uncorrelated orientations per position, totalling $18$ measurements, being this not specific for \acp{RC} with turntables. The \ac{CTIA} addresses \acp{RC} with turntables. In particular, the measurements shall be taken at two different heights, two positions on the turntable, and three uncorrelated orientations, for a total of $12$ measurements. Due to the specific addressing of \acp{RC} with turntables, we opt for a selection of measurements in line with the \ac{CTIA} recommendations, which is described in Section~\ref{MeasCases}.

These measurements are taken using the same stirring sequence as the one used for \ac{DUT} measurements. Therefore, the same number of samples is taken as with the actual \ac{DUT} measurements. Hence, the uncertainty due to the finite number of samples is also considered when evaluating the standard deviation between several measurements. On the other hand, if different polarizations are considered, then this is also considered in these precharacterization measurements. Finally, the lack of spatial uniformity is also considered since the measurements are taken at different positions and orientations within the test volume of the \ac{RC}. The \ac{3GPP} highlights that directive antennas are more prone to detect this lack of spatial uniformity due to their spatial selectivity. Hence, directive antennas should be used for these precharacterization measurements if directive \acp{DUT} are meant to be tested. In summary, the proposed method from \ac{3GPP} and \ac{CTIA}, properly applied, captures the three uncertainty components discussed in Section~\ref{UCComps}.

For convenience, we use the notation from \ac{CTIA}, since it is more detailed. Therefore, we have that the standard deviation of the power transfer function measurements is computed as
\begin{equation}
\sigma_{G_{ref,lin}}=\sqrt{\frac{1}{\left(T_{pre}-1\right)}\sum_{t=1}^{T_{pre}}\left(G_{ref,t,lin}-G_{ref,lin}\right)^2},
\end{equation}
where $T_{pre}$ is the number of uncorrelated power transfer function $G_{ref,t,lin}$ measurements at different positions of the working volume of the chamber as described in \ac{3GPP} and \ac{CTIA} \cite{3GPPRCStandard, CTIAprechar}, and $G_{ref, lin}$ is the average power transfer function over the $T_{pre}$ measurements. $\sigma_{G_{ref,lin}}$ is then converted to dB according to
\begin{equation}
\sigma_{G_{ref,dB}}=\left(\frac{G_{ref,lin}+\sigma_{G_{ref,lin}}}{G_{ref,lin}}\right).
\end{equation}
The measurements follow an approach consistent with the \ac{CTIA} method since it considers \acp{RC} with turntables. Therefore, we describe the positions defined in \ac{CTIA}. In particular, it is stated that $3$ orthogonal orientations of the reference antenna shall be used for each combination of the following values of radii $R$ and heights $Z$:
\begin{itemize}
    \item $R=R_{min}$, where $R_{min}$ is defined as half the minimum distance that the reference antenna can be placed away from the \ac{DUT}. For convenience, we will take this as approximately the center of the turntable.
    \item $R=R_{max}$, where $R_{max}$ is the maximum radius of the test zone, usually defined by the outer radius of the turntable.
    \item $Z=0$, where $0$ refers to the lowest height of the test volume, which should be at least half a wavelength away from the turntable, according to \cite{3GPPRCStandard}.
    \item $Z=Z_{max}$, where $Z_{max}$ refers to the maximum height of the test volume, which should be at least half a wavelength away from any part of the \ac{RC}, be it the ceiling or a stirrer.
\end{itemize}

\subsection{Uncertainty relation with K-factor}
\label{KFUF}
Theoretical analyses of measurement uncertainty in \ac{RC}  can be found in , e.g., \cite{Kildal_RC_KF_formula,RemleyUC1}. In this work, we do not focus on a comparison of theoretical models of uncertainty based on the number of independent samples and $K$-factor. Since we have a constant number of independent samples for all measurements, the only difference in these models can come from the different $ K$ factors. An observation is, however, that the formula from \cite{Kildal_RC_KF_formula}, for all purposes of the considered cases in this study, yields higher uncertainty with larger $K$-factors. We present the average $K$-factor values for each measurement set, as an indicator to quantify the deviation of the generated environment from an ideal Rayleigh channel.

\section{K-factor estimation}
\label{KFEST}
The $K$-factor estimation method that we have implemented for \acp{RC} with turntables is shown further. This method improves existing methods in the literature (see Section~\ref{KFRS}).  This is a relevant contribution since $K$-factor can be used to assess how close to an ideal Rayleigh distribution is the achieved signal in an actual \ac{RC}. Furthermore, estimating the uncertainty of the measurement, considering both the finite sample size and the $K$-factor itself, is a potential area for further future investigation.

\subsection{Proposed method}
\label{KFPM}
We propose to use a combination of the method to estimate $K$-factor in \acp{RC} with turntables presented in \cite{StirUnstir}, with the use of the unbiased estimator of $K$-factor presented in \cite{KFEmulRician}. \cite{StirUnstir} proposes estimating the stirred and unstirred powers for each turntable position, then computing the $K$-factor as the division of the sum of the unstirred powers by the sum of the stirred powers. This $K$-factor value is denoted as average stationary-\ac{LOS} $K$-factor or $\widehat{K}_{av}$, and we denote it as ``literature turntable". On the other hand, \cite{KFEmulRician} proposes an unbiased estimator for the $K$-factor, which is not used in \cite{StirUnstir}.
The method that we propose is the following: we first estimate the $K$-factor for each of the turntable positions using the unbiased estimator from \cite{KFEmulRician}, which is then averaged over all turntable positions, obtaining the average stationary-\ac{LOS} $K$-factor with unbiased estimation or $\widehat{K}_{av,ub}$. All this can be expressed as
\begin{equation}
    \widehat{K}_{tt}=\frac{N_\mathrm{eff}-2}{N_\mathrm{eff}-1}\widehat{K}_{2,tt} -\frac{1}{N_\mathrm{eff}},
\end{equation}
\begin{equation}
         \widehat{K}_{2,tt}=\frac{\widehat{v_{tt}^2}}{\widehat{2\sigma_{tt}^2}},
\end{equation}
\begin{equation}
    \widehat{v_{tt}^2}=\langle S_{21,r}\rangle_{SP_{tt}}^2 + \langle S_{21,i}\rangle_{SP_{tt}}^2,
\end{equation}
\begin{equation}
    \widehat{2\sigma_{tt}^2}=\widehat{\mathrm{Var}}\left[S_{21,r}\right] + \widehat{\mathrm{Var}}\left[S_{21,i}\right],
\end{equation}
where $S_{21}$ is the measured complex transmission coefficient corresponding to a single frequency and containing $SP_{tt}$ samples corresponding to the number of stirrer positions for a given turntable position. Consequently, $\widehat{K}_{2,tt}$ is the biased estimator of the $K$-factor for turntable position $tt$, which ranges from $1$ to $N_{tt}$, being $\widehat{K}_{tt}$ the unbiased estimator. $\widehat{\mathrm{Var}}$ is the unbiased estimator of variance and the brackets $\langle \bullet \rangle_{SP_{tt}}$ denote sample averaging over $SP_{tt}$. Then we have
\begin{equation}
    \widehat{K}_{av,ub}=\frac{1}{N_{tt}}\sum_{1}^{N_{tt}}{\widehat{K}}_{tt}.
\end{equation}
The total number of samples equals $SP_{tt}\times N_{tt}$. For the simulations, we use $\widehat{K}_{av,ub}$, but for the measurements, we also present $\widehat{K}_{av,ub,b}$ which is the frequency averaged $\widehat{K}_{av,ub}$. $K$-factor results are presented in dB, computed by applying $10\log_{10}{(\bullet)}$.

On the other hand, we highlight the relevance of applying an estimation method tailored for \acp{RC} with turntables. We compare the performance of our proposed method and the one from \cite{StirUnstir} with just estimating the $K$-factor ignoring that the data comes from a stirring sequence that includes a turntable. For consistency with the proposed method, we employ the unbiased estimator from \cite{KFEmulRician}. However, both biased and unbiased estimators produce inaccurate results in this case. This is because the presence of a turntable causes the phase of the \ac{LOS} component to fluctuate with the turntable position. Then, if all samples are taken without any particular consideration of the turntable effect and $K$-factor is estimated from them, this can lead to estimating a very low $K$-factor even in presence of a very strong \ac{LOS} component. This is because the phase shift of the \ac{LOS} component across each turntable position is going to be regarded by the estimator as unstirred power \cite{StirUnstir}. We denote this estimator as ``literature no turntable".

\section{Measurement setup}
\subsection{Power transfer function measurement}
We used a Bluetest RTS65 \ac{RC} with dimensions $1945\times2000\times1440$~mm$^3$ (WxHxD) \cite{BluetestCATR}. The measurements were taken using a \ac{VNA} configured to sweep over the lower FR2 band of $24.25$-$29.5$~GHz, with a $10$~MHz step size. This results in $526$ frequency points. \ac{IF} bandwidth was set to $1$~KHz. $600$ complex-valued $S_{21}$ samples were acquired, distributed in $SP_{tt}=24$ positions of the two linear stirrers of the chamber, moving both of them at the same time, and $N_{tt}=25$ turntable positions, covering almost a full circle. For each turntable position, samples are taken for all positions of the linear stirrers. These $600$ samples can safely be assumed to be independent, as shown in previous works using similar measurements by applying the \ac{3GPP} recommended method \cite{ISAP_ours,CorrMatrix} as well as in unpublished data, where a much larger number of samples were taken and they were still independent. Therefore $N_{eff}=600$.

\begin{figure}
\centering
\subfloat[$K$][\label{SU_1}]{\includegraphics[width=0.5\columnwidth]{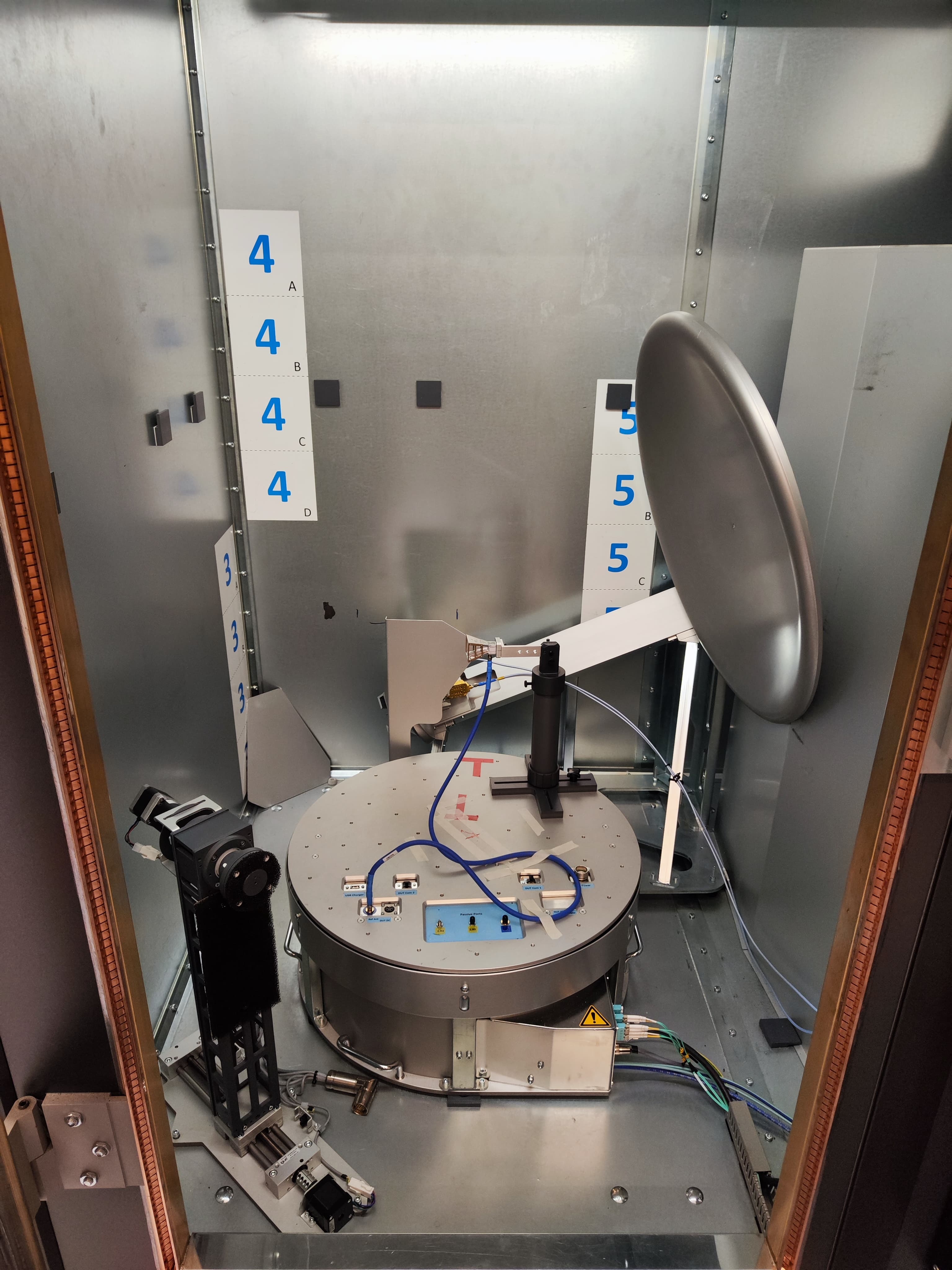}}
\hfil
\subfloat[$K$][\label{SU_2}]{\includegraphics[width=0.5\columnwidth]{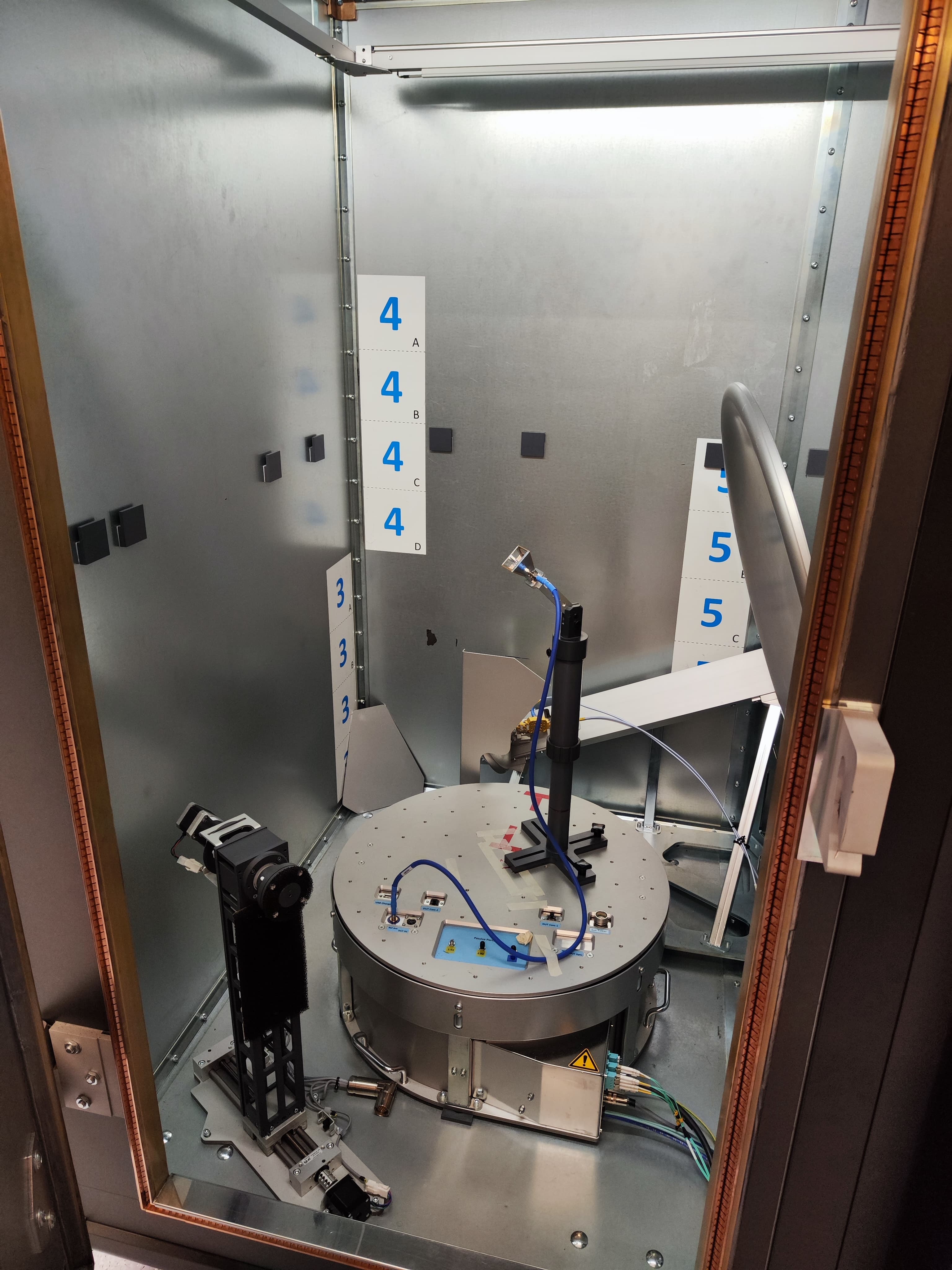}}
\hfil
\caption{Measurement setup. (a) $\mathrm{Z=ZL}$, $\mathrm{R=RM}$,$\mathrm{O=OH}$, and $\mathrm{P=P1}$. (b) $\mathrm{Z=ZH}$, $\mathrm{R=Rm}$,$\mathrm{O=OD}$, and $\mathrm{P=P2}$. The horn antenna can be observed in the turntable attached to a low-loss plastic holder. The structure for holding \ac{mmWave} \acp{DUT} is at the bottom left, on the chamber's floor.}
\label{SU}
\end{figure}

We do not go into detail regarding the performed calibration, since we are just interested in the relative values or variations of the power transfer function. We just say that it is the same calibration the same for all measurements, so they are comparable. The chamber was loaded with the structure that is used for some of the FR2 measurements, as depicted in the left bottom corner in Fig.~\ref{SU}. It was left in the ground of the chamber instead of on the turntable because we were not interested in capturing proximity effects, but rather just keeping the same loading. The antenna used for the precharacterization measurements is a horn antenna with around $14$~dBi gain \cite{DRH50}.

\subsection{Considered cases for precharacterization measurements}
\label{MeasCases}
This paper considers $24$ measurements obtained by combining various configurations of the positioning of the measurement horn antenna: $3$ heights ($\mathrm{Z=ZL,ZM,ZH}$), $2$ positions in the turntable ($\mathrm{R=Rm, RM}$), $2$ antenna orientations, parallel to the turntable ($\mathrm{O=OH}$) and $45^\circ$ upwards inclined ($\mathrm{O=OD}$), and $2$ orthogonal polarizations ($\mathrm{P=P1,P2}$). The heights are referred to the turntable and to the holder of the horn antenna, which corresponds with the height of the horn antenna only for the $\mathrm{O=OH}$ cases, being around $7.5$~cm higher for the $\mathrm{O=OD}$ cases. For the $\mathrm{O=OH}$ cases, we have $\mathrm{ZL=29}$~cm, $\mathrm{ZM=38.5}$~cm, $\mathrm{ZH=48}$~cm, covering the full range of heights of the holder. The turntable positions correspond to placing the horn antenna approximately in the middle ($\mathrm{R=Rm}$) and close to the outer edge ($\mathrm{R=RM}$) of the turntable.

\section{Results}

\subsection{K-factor estimation simulations}
\label{KFRS}
\begin{figure}[!t]
\centering
\includegraphics[width=1\columnwidth]{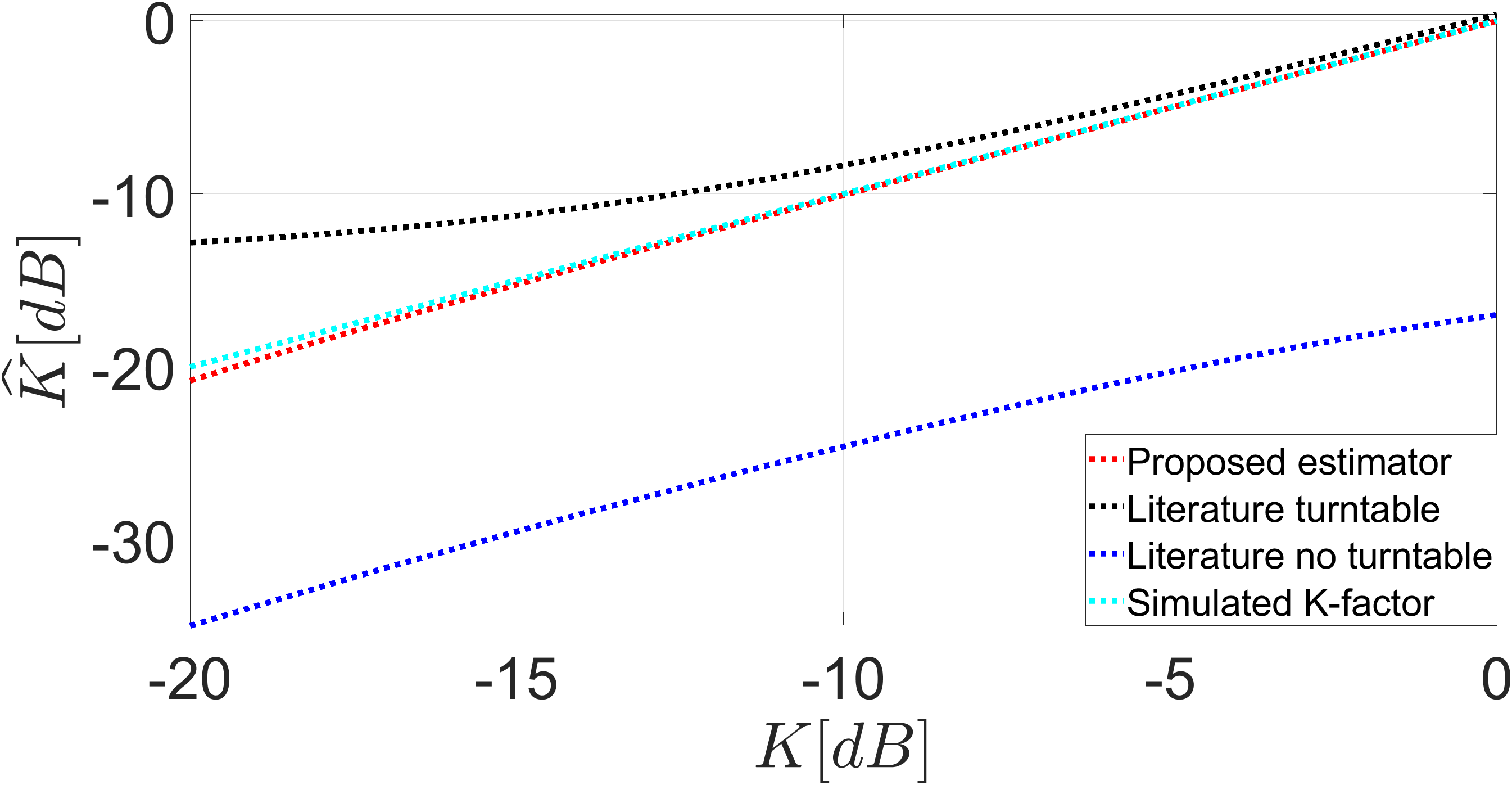}
\caption{Simulation results of the considered $K$-factor estimators.}
\label{F_K_sim}
\end{figure}

\begin{figure}[!t]
\centering
\includegraphics[width=1\columnwidth]{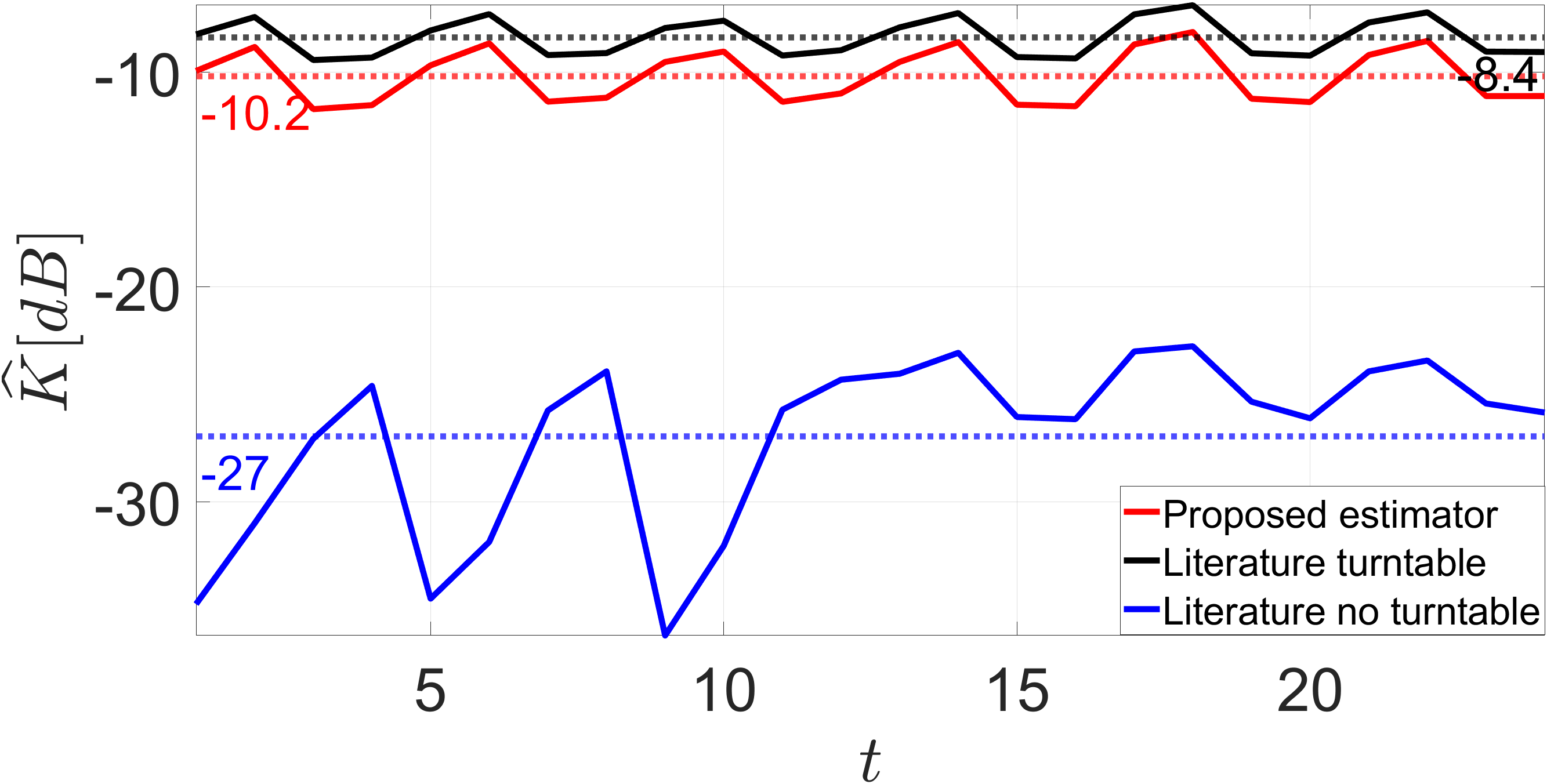}
\caption{Measurement results of the considered $K$-factor estimators. $t$ is the measurement index, ranging from $1$ to $T_{pre}=24$.}
\label{F_K_meas}
\end{figure}
Monte Carlo simulations were run to assess the performance of three considered $K$-factor estimators described in Section~\ref{KFPM}. For this, $N_{eff}=600$ samples were drawn from a Rician distribution with a varying $K$-factor between $-20$ and $0$~dB, with a $1$~dB step size. The samples were divided in $N_{tt}=25$ groups of $SP_{tt}=24$ samples. Each group was subject to a random, uniformly distributed, phase shift, to emulate the phase shift produced when moving from one position in the turntable to another. Then, the $K$-factor was computed according to each estimator. This was repeated $10^{6}$ times, taking the average of the computed $K$-factor values as the final result, expressed in dB. These results are shown in Fig~\ref{F_K_sim}. The proposed estimator outperforms the one from \cite{StirUnstir} (``Literature turntable"), which tends to overestimate the $K$-factor. On the other hand, using the estimator from \cite{KFEmulRician} over all samples, disregarding the turntable effect (``Literature no turntable"), yields a severe underestimation of $K$-factor, rendering this estimator useless. The proposed estimator has some underestimation error along all considered $K$-factor values, being left for future work to explore if this is a bias of the estimator or just a simulation artifact.

On the other hand, in Fig~\ref{F_K_meas} the frequency-averaged estimated $K$-factor values are plotted for each of the $24$ measurements described in Section~\ref{MeasCases}. It can be observed that the estimators behave consistently with simulations. In particular, if we take the average of the estimated $K$-factor values for all measurements, we have that the proposed estimator yields a $K$-factor of $-10.2$~dB, the ``Literature turntable" one yields $-8.4$~dB, and the ``Literature no turntable" one yields $-27$~dB. This is consistent with simulations and indicates that the actual $K$-factor should be close to $-10$~dB, since for a simulated $K$-factor of $-10$~dB, the proposed estimator yields $-10.08$~dB, the ``Literature turntable" one yields $-8.35$~dB, and the ``Literature no turntable" one yields $-24$~dB. As for the $K$-factor value itself, we refrain from qualifying it as better or worse, since this work is not about the comparison of different \acp{RC}. Just for reference, it is on par with the $K$-factor of a \ac{VIRC} \cite{ZaherVIRCKF} and a bit lower (better) than the obtained in \cite{StirUnstir}, although there is a large difference in the frequency range. This $K$-factor value is sufficiently low to pass a \ac{GoF} test for Rayleigh distribution, as shown in \cite{ZaherVIRCKF,EuCAP2024_ours}.

\subsection{Precharacterization measurements}
\begin{table}[!t]
\caption{}
\centering
\includegraphics[width=1\columnwidth]{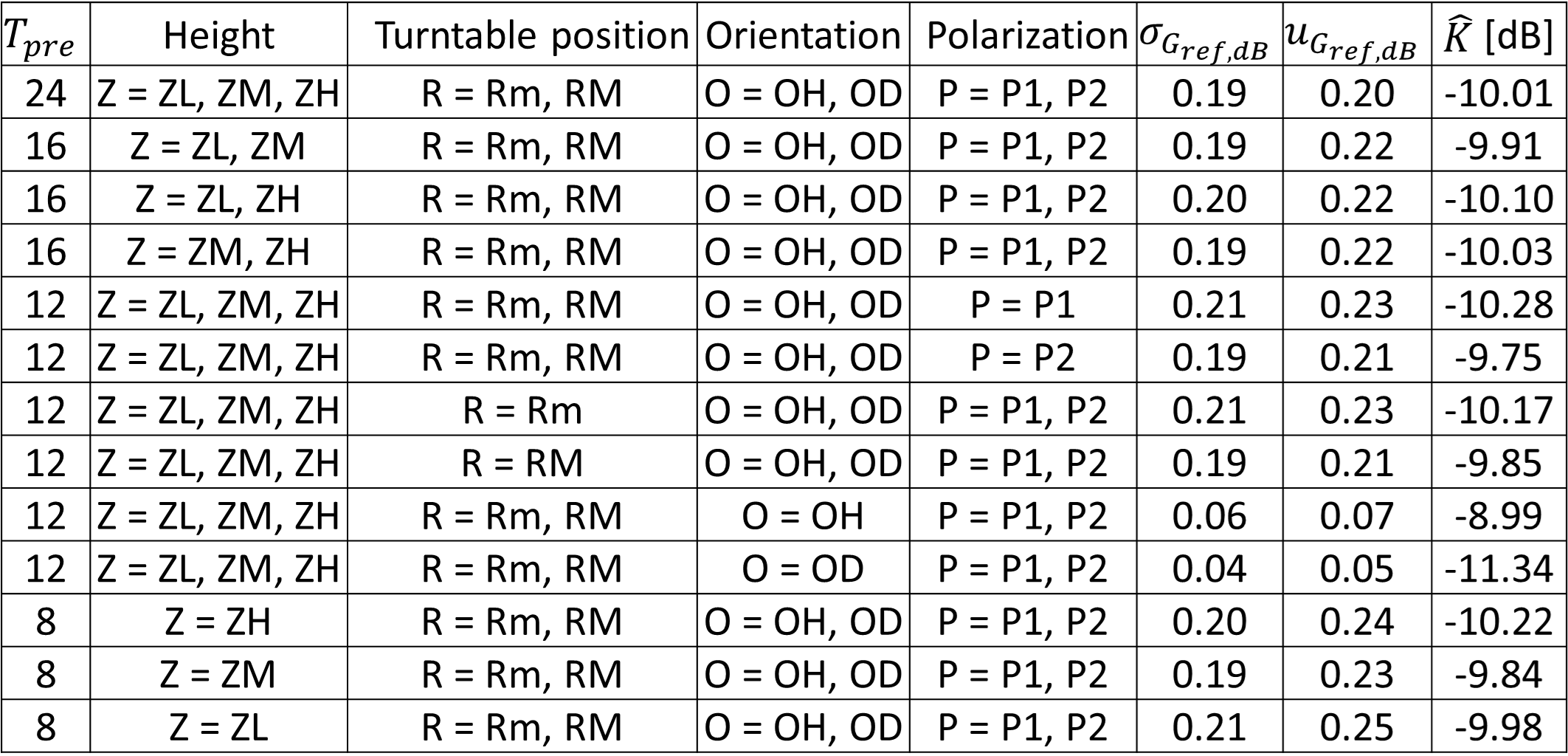}
\label{Table_GU}
\end{table}

We analyze $\sigma_{G_{ref,dB}}$ for different sets of measurements. First, we take all $24$ measurements ($T_{pre}=24$), then we remove one of the $3$ heights, thus having $T_{pre}=16$, then we consider one polarization, position on the turntable and orientation at a time, which results in $T_{pre}=12$, and finally we take just one height, thus having $T_{pre}=8$. Since we vary $T_{pre}$, we need to take into account the statistic effects of estimating $\sigma_{G_{ref,dB}}$ from a different number of measurements, thus enabling comparison of the results across different $T_{pre}$. Therefore, following \cite{CTIAMU}, we apply a correction factor $K_p$, thus having $u_{G_{ref,lin}}=\sigma_{G_{ref,lin}}\times K_p$. $K_p$ results from dividing the coverage factor for a $95\%$ \ac{CI} for $T_{pre}-1$ degrees of freedom by the coverage factor for a $95\%$ \ac{CI} for $\infty$ degrees of freedom ($1.96$). These results are summarized in Table~\ref{Table_GU}, which also includes the $T_{pre}$-averaged estimated $K$-factor with our proposed method. Since it is required that all $T_{pre}$ measurements are uncorrelated, the Pearson correlation coefficients were computed for all pairs of sets of $600$ samples, at each frequency. They were then compared to the same threshold used in \cite{CorrMatrix}, being lower in all cases, thus concluding that all $24$ measurements are uncorrelated.

Moreover, an \ac{ANOVA} has been performed, taking the frequency-averaged power transfer function samples ($N_{eff}=600$ per measurement) of the $24$ measurements as inputs. The linear regression model that is fitted is $Y\sim1+\mathrm{P}+\mathrm{R}+\mathrm{Z}+\mathrm{O}$, being $Y$ the power transfer function, from where coefficients for each variable are obtained. The results of the \ac{ANOVA} are shown in Table~\ref{Table_ANOVA}.
\begin{table}[!t]
\caption{ANOVA and Linear Regression Coefficients}
\centering
\includegraphics[width=1\columnwidth]{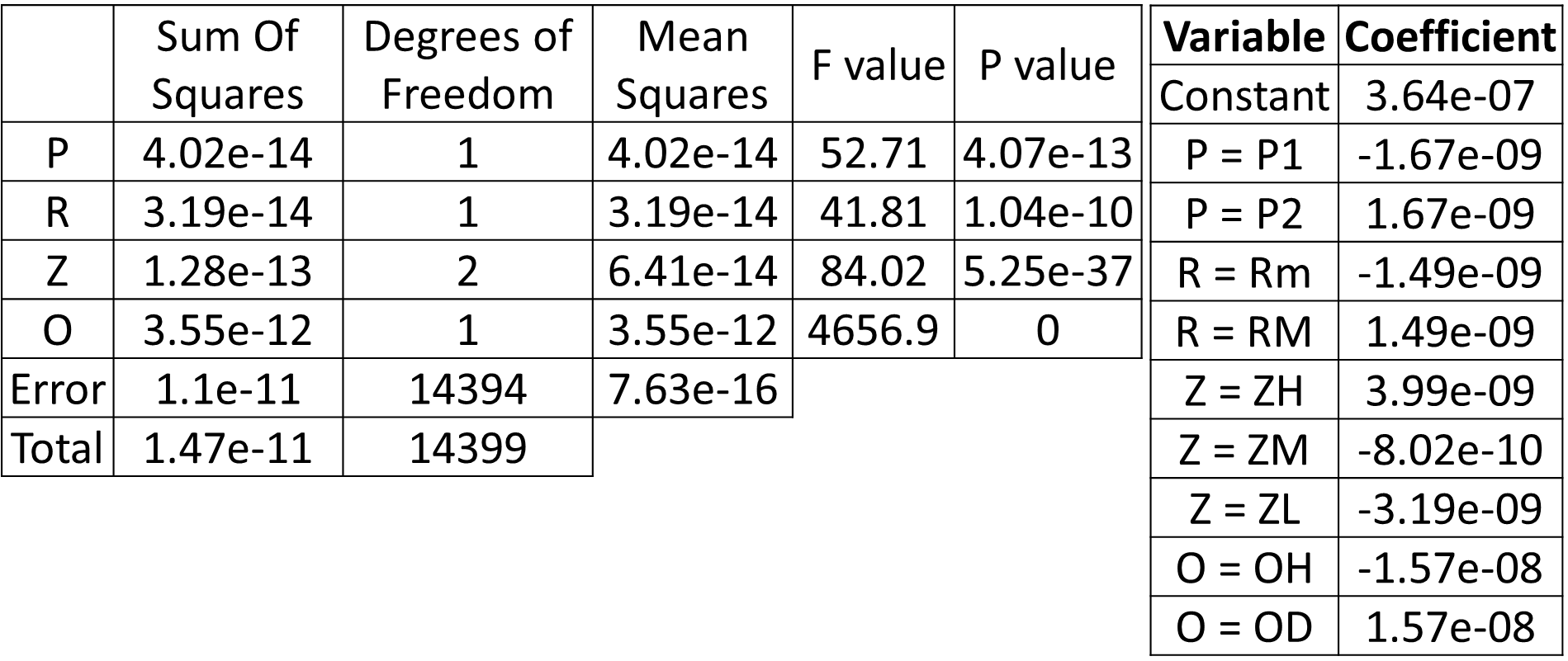}
\label{Table_ANOVA}
\end{table}

Several highly relevant observations can be made. Firstly, all variables significantly affect the variability of the power transfer function, as indicated by very low p-values. Linear regression coefficients for each dummy variable show the effect on the mean of $Y$ when other variables are held constant. Although the direct comparison of coefficients is complex due to the three possible values for height and significant interactions found in the \ac{ANOVA}, the coefficients for orientation $\mathrm{O}$ are roughly an order of magnitude higher. This suggests that varying $\mathrm{O}$ has a substantial impact on $Y$'s variability, leading to a higher $\sigma_{G_{ref,dB}}$.

Table~\ref{Table_GU} confirms this: with $\mathrm{O}$ constant at $\mathrm{OH}$ or $\mathrm{OD}$, $\sigma_{G_{ref,dB}}$ and $u_{G_{ref,dB}}$ are significantly lower. When $\mathrm{O}$ varies between $\mathrm{OH}$ and $\mathrm{OD}$, $\sigma_{G_{ref,dB}}$ and $u_{G_{ref,dB}}$ values are similar, though $u_{G_{ref,dB}}$ increases slightly with smaller $T_{pre}$. If a slightly higher $u_{G_{ref,dB}}$ is acceptable for the uncertainty budget, varying $\mathrm{O}$ allows a substantial reduction in precharacterization measurement time (up to one-third for fixed height cases) with minimal impact on $u_{G_{ref,dB}}$, representing a favorable trade-off.

The pronounced effect of orientation on variability may indicate a non-uniform elevation distribution of \acp{AoA}, which can happen under certain \ac{RC} geometries, as explored in \cite{TTstirring_8}. Regarding $K$-factor values, no clear relation with $u_{G_{ref,dB}}$ is observed, suggesting additional factors beyond $K$-factor are at play in precharacterization measurements. The higher $K$-factor for $\mathrm{OH}$ cases compared to $\mathrm{OD}$ cases implies more unstirred power or \ac{LOS} components in the azimuthal plane (captured by $\mathrm{OH}$). This could be due to the presence of a ceiling stirrer and the use of a directive antenna, which might enhance mode stirring effectiveness when oriented towards the stirrer, thus reducing unstirred power and $K$-factor.

\section{Conclusions}
In this study, we analyzed the uncertainty in the power transfer function of an \ac{RC}, emphasizing the impact of precharacterization measurement variables. We introduced a new $K$-factor estimator for \acp{RC} using turntables, which performs better than existing methods, as demonstrated by Monte Carlo simulations.

Our findings reveal that, among the various precharacterization variables, the antenna orientation is the most significant factor for the \ac{mmWave} directive antenna setup. All variables substantially affect power transfer function variability, but optimal results are achieved by varying the antenna orientation rather than just increasing the number of measurements. This conclusion is specific to the \ac{RC} design and may vary with different designs or antenna placements; for instance, polarization might become more relevant in cases of significant imbalance.

We also observed potential non-uniform \ac{AoA} distribution in the elevation plane due to the pronounced effect of antenna orientation on power transfer function variability. Additionally, a greater mode stirring effect was noted when deviating from the azimuthal plane, as indicated by the higher $K$-factor. No significant relationship was found between the $K$-factor and $u_{G_{ref,dB}}$, suggesting that future research should explore whether uncertainty formulas based on limited samples and $K$-factor are reliable for \ac{mmWave} and directive antennas, or if other factors are more influential.

\section*{ACKNOWLEDGEMENT}
This work is within the Horizon 2020 MSCA-ITN 5VC project, supported by the EU’s research and innovation program under grant agreement No. 955629. Funding from the ELLIIT strategic research environment (https://elliit.se/) is also kindly acknowledged.

\bibliographystyle{IEEEtran}

\bibliography{References}

\end{document}